\documentstyle[11pt,psfig,twoside]{article}

\begin{document}

\newcommand{\dd}{\,{\rm d}}
\newcommand{\ie}{{\it i.e.},\,}
\newcommand{\etal}{{\it et al.\ }}
\newcommand{\eg}{{\it e.g.},\,}
\newcommand{\cf}{{\it cf.\ }}
\newcommand{\vs}{{\it vs.\ }}
\newcommand{\zdot}{\makebox[0pt][l]{.}}
\newcommand{\up}[1]{\ifmmode^{\rm #1}\else$^{\rm #1}$\fi}
\newcommand{\dn}[1]{\ifmmode_{\rm #1}\else$_{\rm #1}$\fi}
\newcommand{\upd}{\up{d}}
\newcommand{\uph}{\up{h}}
\newcommand{\upm}{\up{m}}
\newcommand{\ups}{\up{s}}
\newcommand{\arcd}{\ifmmode^{\circ}\else$^{\circ}$\fi}
\newcommand{\arcm}{\ifmmode{'}\else$'$\fi}
\newcommand{\arcs}{\ifmmode{''}\else$''$\fi}
\newcommand{\MS}{{\rm M}\ifmmode_{\odot}\else$_{\odot}$\fi}
\newcommand{\RS}{{\rm R}\ifmmode_{\odot}\else$_{\odot}$\fi}
\newcommand{\LS}{{\rm L}\ifmmode_{\odot}\else$_{\odot}$\fi}

\newcommand{\Abstract}[2]{{\footnotesize\begin{center}ABSTRACT\end{center}
\vspace{1mm}\par#1\par
\noindent
{~}{\it #2}}}

\newcommand{\TabCap}[2]{\begin{center}\parbox[t]{#1}{\begin{center}
  \small {\spaceskip 2pt plus 1pt minus 1pt T a b l e}
  \refstepcounter{table}\thetable \\[2mm]
  \footnotesize #2 \end{center}}\end{center}}

\newcommand{\TableSep}[2]{\begin{table}[p]\vspace{#1}
\TabCap{#2}\end{table}}

\newcommand{\FigCap}[1]{\footnotesize\par\noindent Fig.\  %
  \refstepcounter{figure}\thefigure. #1\par}

\newcommand{\TableFont}{\footnotesize}
\newcommand{\TableFontIt}{\ttit}
\newcommand{\SetTableFont}[1]{\renewcommand{\TableFont}{#1}}

\newcommand{\MakeTable}[4]{\begin{table}[htb]\TabCap{#2}{#3}
  \begin{center} \TableFont \begin{tabular}{#1} #4 
  \end{tabular}\end{center}\end{table}}

\newcommand{\MakeTableSep}[4]{\begin{table}[p]\TabCap{#2}{#3}
  \begin{center} \TableFont \begin{tabular}{#1} #4 
  \end{tabular}\end{center}\end{table}}

\newenvironment{references}%
{
\footnotesize \frenchspacing
\renewcommand{\thesection}{}
\renewcommand{\in}{{\rm in }}
\renewcommand{\AA}{Astron.\ Astrophys.}
\newcommand{\AAS}{Astron.~Astrophys.~Suppl.~Ser.}
\newcommand{\ApJ}{Astrophys.\ J.}
\newcommand{\ApJS}{Astrophys.\ J.~Suppl.~Ser.}
\newcommand{\ApJL}{Astrophys.\ J.~Letters}
\newcommand{\AJ}{Astron.\ J.}
\newcommand{\IBVS}{IBVS}
\newcommand{\PASP}{P.A.S.P.}
\newcommand{\Acta}{Acta Astron.}
\newcommand{\MNRAS}{MNRAS}
\renewcommand{\and}{{\rm and }}
\section{{\rm REFERENCES}}
\sloppy \hyphenpenalty10000
\begin{list}{}{\leftmargin1cm\listparindent-1cm
\itemindent\listparindent\parsep0pt\itemsep0pt}}%
{\end{list}\vspace{2mm}}

\def\TYLDA{~}
\newlength{\DW}
\settowidth{\DW}{0}
\newcommand{\dw}{\hspace{\DW}}

\newcommand{\refitem}[5]{\item[]{#1} #2%
\def\REFARG{#3}\ifx\REFARG\TYLDA\else, {\it#3}\fi
\def\REFARG{#4}\ifx\REFARG\TYLDA\else, {\bf#4}\fi
\def\REFARG{#5}\ifx\REFARG\TYLDA\else, {#5}\fi.}

\newcommand{\Section}[1]{\section{#1}}
\newcommand{\Subsection}[1]{\subsection{#1}}
\newcommand{\Acknow}[1]{\par\vspace{5mm}{\bf Acknowledgments.} #1}
\pagestyle{myheadings}

\def\thefootnote{\fnsymbol{footnote}}
%%%%%%%%%%%%%%%%%%%%%%%%%%%%%%%%%%%%%%%%%%%%%%%%%%%%%%%%%%%%%%%%%%%%%%%%%%%%

\begin{center}
{\Large\bf Period Changes in Galactic Classical Cepheids.
Slow Evolution of Long-period Cepheids}
\vskip1cm
{\bf
P~a~w~e~\l\ ~~P~i~e~t~r~u~k~o~w~i~c~z}
\vskip3mm
{Copernicus Astronomical Center, Bartycka 18, 00-716 Warszawa, Poland\\
e-mail: pietruk@camk.edu.pl}
\end{center}

\Abstract{
We compared period changes derived from $O-C$ diagrams for 63
classical Cepheids from our Galaxy with model calculations.
We found that for Cepheids with ${\rm log ~ P > 1.0}$ the observed
changes are smaller than predicted values, except variable
SZ Cas. However some of the first overtone Cepheids, particularly
EU Tau and Polaris, change its period much faster than it follows
from theory. Summary of the known data on the period changes in
Cepheids from the Galaxy and from the Magellanic Clouds (previous
papers) leads to conclusion that none of the 999 Cepheid
is undergoing the first crossing of the instability strip.
Also the observed period changes for long-period Cepheids
are a few times slower than predicted by the models.
These results imply that much larger fraction of helium is
burnt in the Cepheid stage than it is predicted by models.
}{Stars: evolution -- Cepheids -- Galaxy}

\Section{Introduction}%1

Classical Cepheids belong to important objects in modern
astrophysics. Their Period-Luminosity relation is a popular
method for estimating extragalactic distances. These variable stars
also provide tests for models of the structure, evolution
and pulsation of stars in general.

Cepheids are massive Population I stars crossing the instability strip
in the Hertzsprung--Russell diagram at the effective temperature 
${\rm \log T_{eff} \approx 3.8}$. Most of them are
undergoing core helium burning. The small fraction of them
may be rapidly crossing the strip
for the first time when stars just left the Main Sequence
and are going redwards. This is a result of the evolution
on the thermal time scale. The following crossings, II and III,
occur during the blue loop phase and take much more time.

When a star crosses the instability strip its pulsation period
changes. Even for massive objects, whose evolutionary time-scales
are relatively and hence the expected rates of period changes are
relatively high, a long time interval is needed to find measurable
changes. Several Cepheids in our Galaxy have been observed
for almost two centuries. Recently Berdnikov {\it et al.}
(2000a) compiled the observations of $ \delta $ Cep, $ \eta $ Aql
and $ \zeta $ Gem, all showing significant period trends.
For very well-known Polaris, beside a strange decrease of
amplitude, the secular period changes are also reported
(Kamper and Fernie 1998, Evans {\it et al.} 2002). Another galactic
Cepheid, Y Oph, also reveals both period and amplitude variations
(Fernie 1990, Fernie {\it et al.} 1995).
Regular photometric monitoring of galactic Cepheids from both
hemispheres was undertaken by Berdnikov (1995,2001).
Also Bersier (2002) analysed a significant amount of photometric
and radial velocity data for 62 Cepheids to find possible binarity
or period changes. Earlier, Turner (1998) published data on period
changes of 137 northern Cepheids and compared them with the
theory predictions. Saitou (1989) tried to discover effects
of metal abundance on the evolutionary period changes and
concluded that there is only a marginally dependence. However,
both autors did not take into account the influence of errors
and used an oversimplified model for comparison with their data.

Recently significant period changes for 67 (mainly northern)
Cepheids were listed by Berdnikov and Ignatova (2000b)
as a result of the analysis of $O-C$ diagrams for 230 objects.
Using these data we compared observed period changes with
the predictions from the recent stellar evolutionary models
by Bono {\it et al.} (2000). In the second part of this
paper we discuss a confrontation
between the calculated and observed to date changes for
more numerous sample of Cepheids. We add much larger
lists of analysed Cepheids in the Large and Small Magellanic
Cloud published by Pietrukowicz (2001, Paper I) and
Pietrukowicz (2002, Paper II), respectively.

\Section{Period changes in Galactic Cepheids}%2

\Subsection{Observational Data}

Among 67 objects listed by Berdnikov and Ignatova (2000b)
we found 61 fundamental mode and three first overtone classical
Cepheids. Two other variables (BL Her and UY Eri) belong to
Population II Cepheids what was checked with a list
of this type of variables prepared by Harris (1985).
Another star, V473 Lyr, is a short-period pulsating variable star,
probably with Blazhko effect. Its identification with a Cepheid
is still not sure (Koen 2001). These three objects were not
included in the present analysis.

For each Cepheid Berdnikov and Ignatova (2000b) give parabolic
elements $M_0$, $P$ and $q \pm \sigma _q$ in the equation
$$
M = M_0 + P \cdot E + q \cdot E^2
\eqno(1)
$$
where $E$ is the epoch number. We calculate the rate of the observed
period change as
$$
r \equiv \frac{2q}{P^3} = \frac{\dot P}{{P}^2}
\eqno(2)
$$
and its uncertainty as
$$
\sigma _r \equiv \frac{2\sigma _q}{P^3}
\eqno(3)
$$
The scaling factor is chosen (as in the previous papers)
so that all model result and observational points can be
clearly displayed in one figure.

For one long-period Cepheid, AQ Pup, we decided to construct a new
$O-C$ diagram (Fig. 1) because period variations
and large relative error of $q$ derived from looked strange.
Berdnikov and Ignatova (2000b). We used new numerous
data (75 observations) made in 2001 by ASAS (the All Sky
Automated Survey, Pojma\'nski 2000), which were kindly
provided before publication. Table 1 lists the $O-C$ data for AQ Pup
maxima collected from the literature. Our estimation of
the period does not reveal any significant variations.
The least squares linear fit gives the following ephemeris:
$$
{\rm HJD}_{max} = (2451933.653 \pm 0.010) + (30.10 \pm 0.18) \times E
\eqno(4)
$$

\begin{table}
\begin{center}
\caption{\small Times of maxima of AQ Pup}

\vspace{0.4cm}

\begin{tabular}{rcrl}
\hline
$E$ & HJD$_{max}$ & $O-C$ & Source of Data
\\
\hline
-808 & 27609.150 &  0.014 & GCVS2 (1958)\\
-338 & 41760.630 &  2.332 & Madore (1975) \\
-230 & 45010.802 &  1.207 & Moffett {\it et al.} (1984)  \\
 -70 & 49824.365 & -1.966 & Berdnikov {\it et al.} (1995) \\
   0 & 51933.653 &  0.000 & ASAS, in prep. \\
\hline
\end{tabular}
\end{center}
\end{table}

\Subsection{Comparison with Evolutionary Models}

The most important properties of all evolutionary models were
described in Paper I. A recent theoretical
survey of Cepheids' characteristics for a number of
evolutionary models was published by Bono {\it et al.} (2000).
For comparison we chose a set of models for
the chemical composition parameters
${\rm Z=0.022, Y=0.289}$, representative for the Galaxy.
The evolutionary tracks were constructed by adopting
the ZAMS mass range ${\rm 3 - 13 M_\odot }$.
Using a linear nonadiabatic pulsation code, we calculated values
of the period changes for Bono {\it et al.} (2000)
models in about twenty points of time for each crossing through
the instability strip.

Fig. 2 presents a comparison between rates of period changes
for the fundamental mode Cepheids and the models of Bono {\it et al.}
(2000). Theoretical predictions for the three instability crossings
are clearly separated. None of the star appears to undergo the rapid first
crossing. Recently Luck {\it et al.} (2001) considered SV Vul
as a possible first crossing Cepheid. Basing on spectroscopic
observations, they suggested that its surface abundances reflect
its original composition because it does not exhibit any dredge-up
effect. Unfortunately, they ignored
its strongly significant negative period change,
which was derived by Berdnikov and Ignatova (2000b) and earlier
also by Turner {\it et al.} (1999). For most Cepheids with
${\rm log ~ P > 1.0}$ the observed period change rates are lower
than models imply. Only SZ Cas, with ${\rm P=13.621}$d,
apears to be rapidly changing its period,
a few times faster than other variables within the
${\rm 1.0 < log ~ P < 1.5}$ range. On the other hand the period
changes are too slow to be explained as the first crossing evolution.

Another interesting point to notice is that, as presented in Fig. 3,
the number of Cepheids with positive period changes (crossing III)
is larger than a number of Cepheids with negative period changes
(crossing II). This fact is also easy recognizable
in the data samples published by Turner (1998) and
Turner {\it et al.} (1999).

Fig. 4, displays the observed period changes for three first
overtone pulsators and the corresponding model values.
Two stars, EU Tau and Polaris, show much higher rate of period
change than predicted for crossing III but also much lower than
for the first crossing. This is especially clear for long-monitored
Polaris. In fact, the discrepancy between the observed and
calculated rates of period changes for the first crossing are
larger than Fig. 4 may suggest. It is because Polaris is apparently
closer to the blue edge of the instability strip (as argued
by Evans {\it et al.} 2002) and would be located well
above the upper limit of Fig. 4. Therefore, we conclude
that Polaris cannot be classified as the first crossing Cepheid.
It is important that comparable large period changes were also
reported for several SMC overtone Cepheids (Paper II).
Evans {\it et al.} (2002) claim that the changes for this type
of variables can be explained by non-evolutionary effects
such as nonlinear effects leading to mode coupling.

\Subsection{Conclusions}

We compared determinations of the rates of period change for
63 galactic Cepheids with model calculations. The observed
changes are significant. All but two Cepheids have
relative error of the rate between 1.5 and 25 percent.
However Berdnikov and Ignatova (2000b) wrote that they analysed
over 200 $O-C$ diagrams. It is likely that none of the
Cepheid from entire their list is undergoing the rapid first
crossing. We also note that we do not know the upper limits on possible
period changes and we were not able to include all sample
for the comparison. Basing on data for 63 objects we showed that,
just as it was the case of Cepheids in the Magellanic Clouds
(Paper I and Paper II), galactic Cepheids with longest periods,
${\rm log ~ P > 1.0}$, appear to evolve slower than it was predicted
by models calculated with metallicity representative for the Galaxy.
It would be interesting to compare period changes for more numerous
sample of stars. A regular photometric monitoring
of Cepheids from both hemispheres should increase the number
of variables with significant period changes in the following years.

\Section{General Conclusion Regarding Period Changes
in the LMC, SMC and Galactic Cepheids}%3

In the previous work we determined the period changes for 378 and
557 Cepheids in the LMC (Paper I) and SMC (Paper II), respectively.
Adding 64 galactic classical Cepheids to those samples give
a total number of 999 Cepheids, for which the rate of period
change was analysed. Certainly none of the Cepheid is undergoing
the first crossing, which takes place on the thermal time scale.
Theory predicts the first crossing time to be a few tens times
shorter than time for crossings II and III, as displayed in Fig. 5.
This ratio is similar for three metallicities, ${\rm Z=0.004, Z=0.01,
Z=0.02}$, representative of the SMC, LMC and the Galaxy, respectively.
The increase is due to an earlier helium ignition which at higher
masses takes place already during Hertzsprung gap passage. The
ratios of crossing times are anyhow too large to explain observations.

Another very important conclusion from our analysis concerns small
values of period change rates for long-period Cepheids.
We found that for Cepheids with ${\rm log ~ P > 1.0}$ the changes
are generally a few times slower than in the models. It is
Interesting that several Cepheids with long periods appear to evolve
much faster than the others, {\it e.g.} HV 829 in the SMC or SZ Cas.
Therefore regular period monitoring of these variables is
particularly necessary.

The comparison clearly shows that, for massive stars, the evolution
through the instability phase is slower than predicted by models.
This means that substantially larger amount of helium is burnt
in this phase. As presented in Fig. 6,
the ratio of duration of crossings II and III to the total
He burning stage, derived from Bono {\it et al.} (2000),
is just 5-8 percent for 6 $M_\odot$ and only 0.4-0.7 percent for
10 $M_\odot$ star, depending on the metallicity.
The real problem begins for stars with masses larger than about
7 $M_\odot$., {\it i.e.} for Cepheids with ${\rm log ~ P > 1.0}$.
Simple calculations indicate that ten times longer blue loop crossing
times ({\it i.e.} ten times lower rates of period change)
would also mean approximately ten times lower ratio
of crossing I Cepheids to Cepheids in crossings II and III.
We believe that the predicted rates of the first crossing seem
robust as they are much less affected by the uncertainty
in the convection treatment and overshooting efficiency
than the models in the loop phase.

There is also another observational evidence for a slower evolution
of massive stars while they cross the instability strip.
Alcock {\it et al.} (2000) present a comprehensive analysis
of color-magnitude diagrams for the brightest stars in the LMC.
They determined the number of nonvariable red supergiants, blue
supergiants and Cepheids, all with magnitude $V < 16.0$, as $r=2064$,
$b=805$ and $c=280$, respectively. Making a 10\% correction to $r$ for
foreground Galactic disk stars, they found $c/(b+r)$ ratio
equal to 0.105. This value corresponds to the fraction of core helium
burnt in the strip and is in a very good agreement with the value
implied by our determination of the period changes in Cepheids.

\Acknow{

I would like to thank Dr. G. Pojma\'nski for providing 
photometric data on AQ Pup before publishing,
and Dr. P. Moskalik for an access to the list of overtone pulsators.
I am greatful to Dr. W. Dziembowski for providing the pulsation code
and important discussions..
It is also a pleasure to thank Dr. B. Paczy\'nski for useful remarks.
}

\begin{figure}[htb]
\hglue-0.5cm\psfig{figure=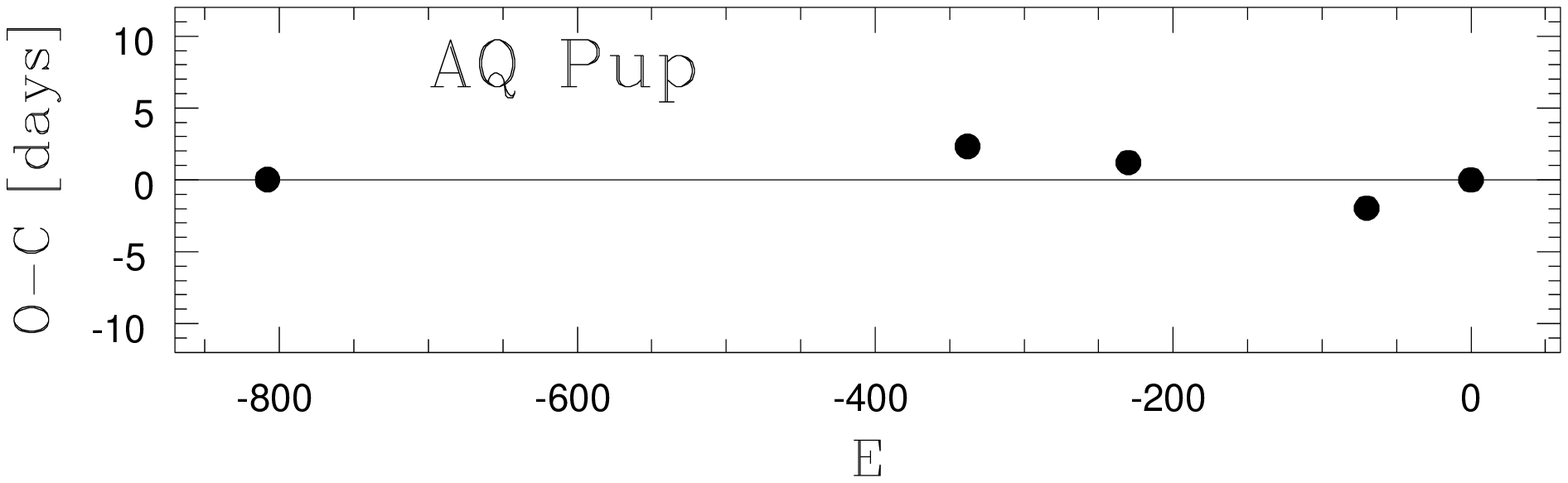,bbllx=0pt,bblly=0pt,bburx=590pt,bbury=720pt,width=12.5cm,
clip=,angle=0}
\vspace*{3pt}
\FigCap{
$O-C$ diagram for Cepheid AQ Pup. The period appears to be constant.
}
\end{figure}

\begin{figure}[htb]
\hglue-0.5cm\psfig{figure=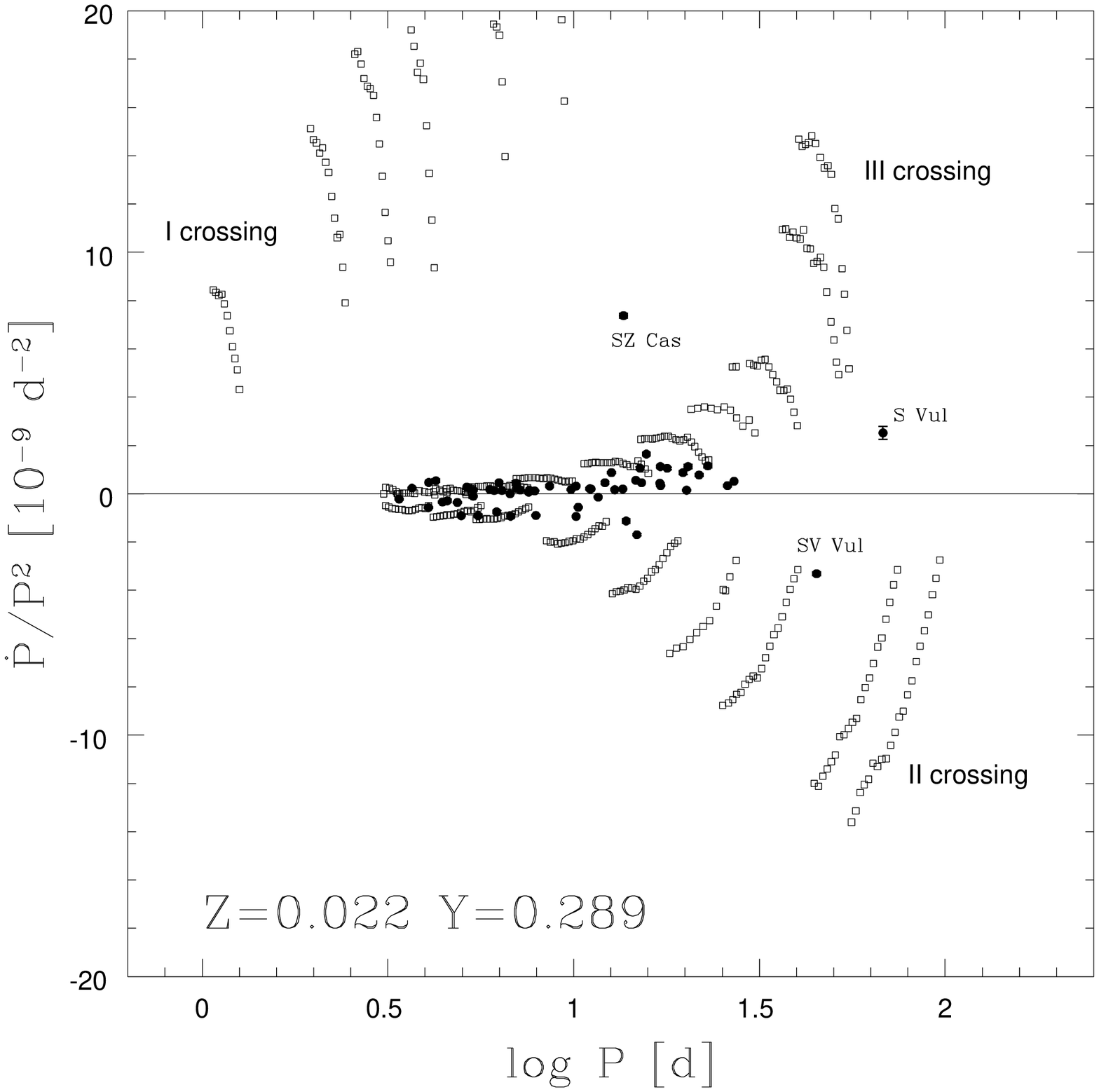,bbllx=0pt,bblly=0pt,bburx=590pt,bbury=720pt,width=12.5cm,
clip=,angle=0}
\vspace*{3pt}
\FigCap{
Comparison between the observed significant period changes for
60 fundamental mode galactic Cepheids and changes predicted
by Bono {\it et al.} (2000). Notice the lack of first crossing
Cepheids. The changes for Cepheids with ${\rm log ~ P > 1.0}$ are
generally slower than predicted, except variable SZ Cas.
}
\end{figure}

\begin{figure}[htb]
\hglue-0.5cm\psfig{figure=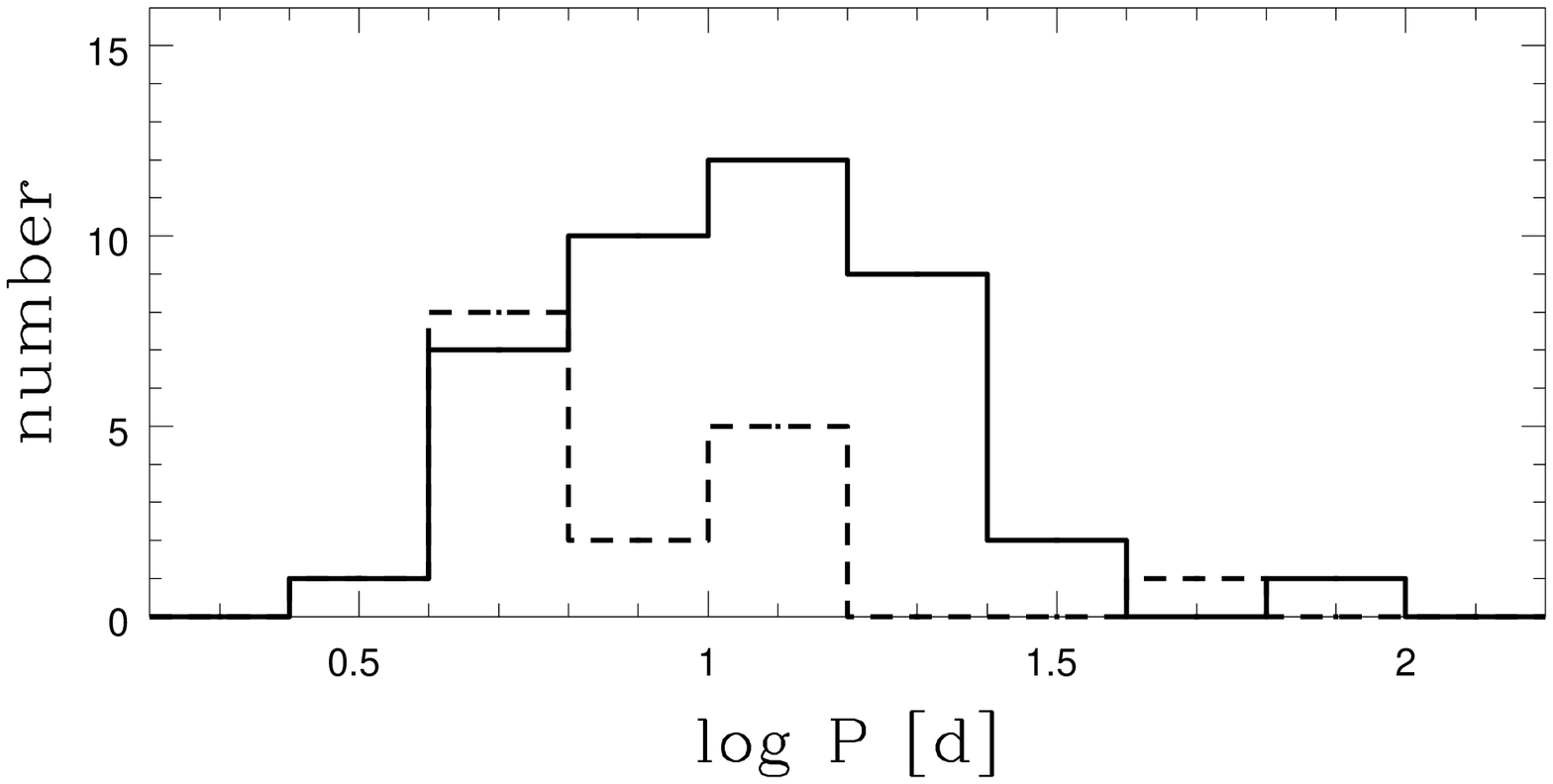,bbllx=0pt,bblly=0pt,bburx=590pt,bbury=720pt,width=12.5cm,
clip=,angle=0}
\vspace*{3pt} 
\FigCap{
Histograms for positive (solid line) and negative (dashed line)
significant period changes. The number of Cepheids with positive
changes (corresponding to crossing III) is generally larger.
}
\end{figure}

\begin{figure}[htb]
\hglue-0.5cm\psfig{figure=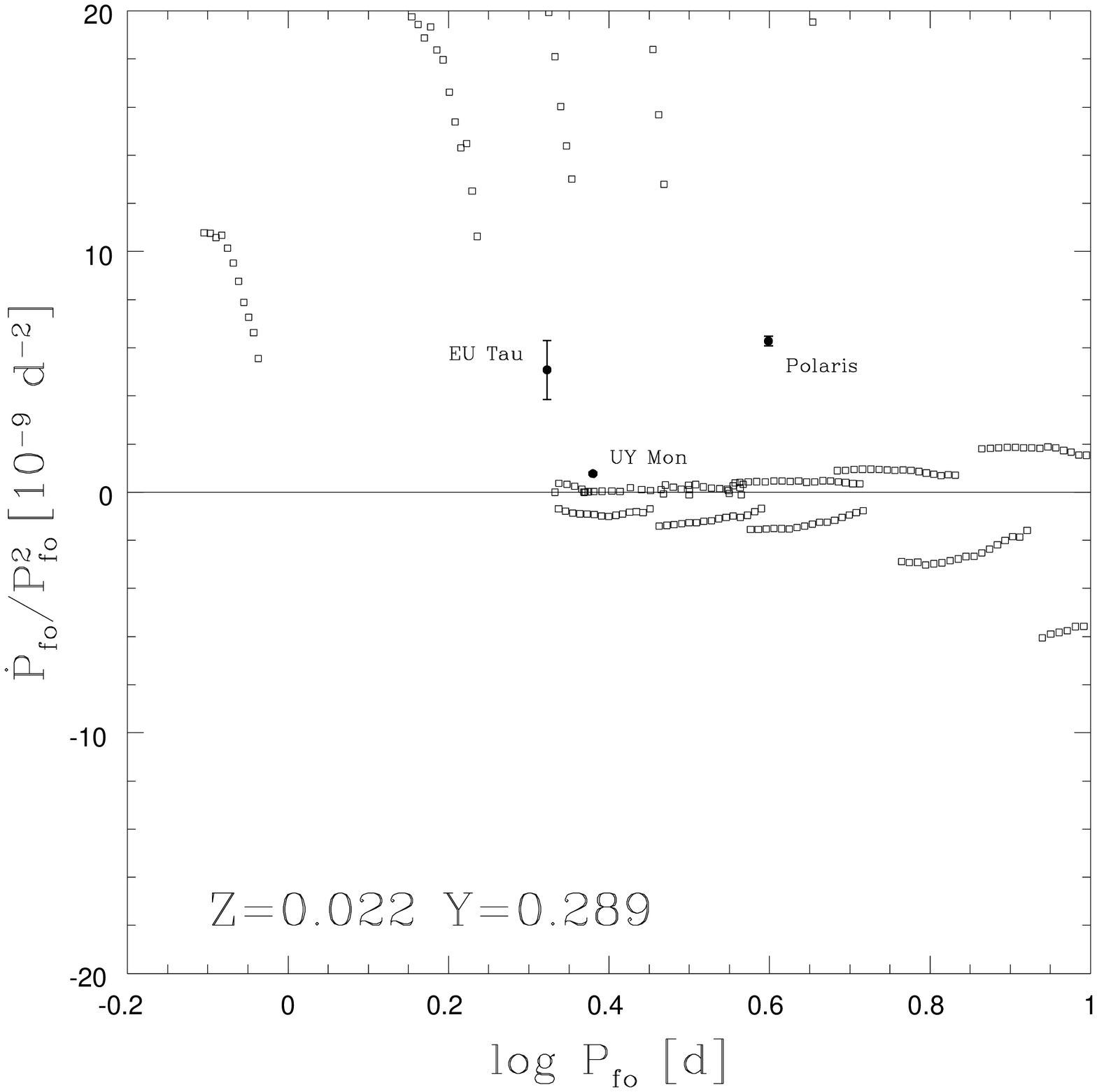,bbllx=0pt,bblly=0pt,bburx=590pt,bbury=720pt,width=12.5cm,
clip=,angle=0}
\vspace*{3pt}
\FigCap{
The observed period changes for three first overtone galactic Cepheids
are compared with models given by Bono {\it et al.} (2000).
}
\end{figure}

\begin{figure}[htb]
\hglue-0.5cm\psfig{figure=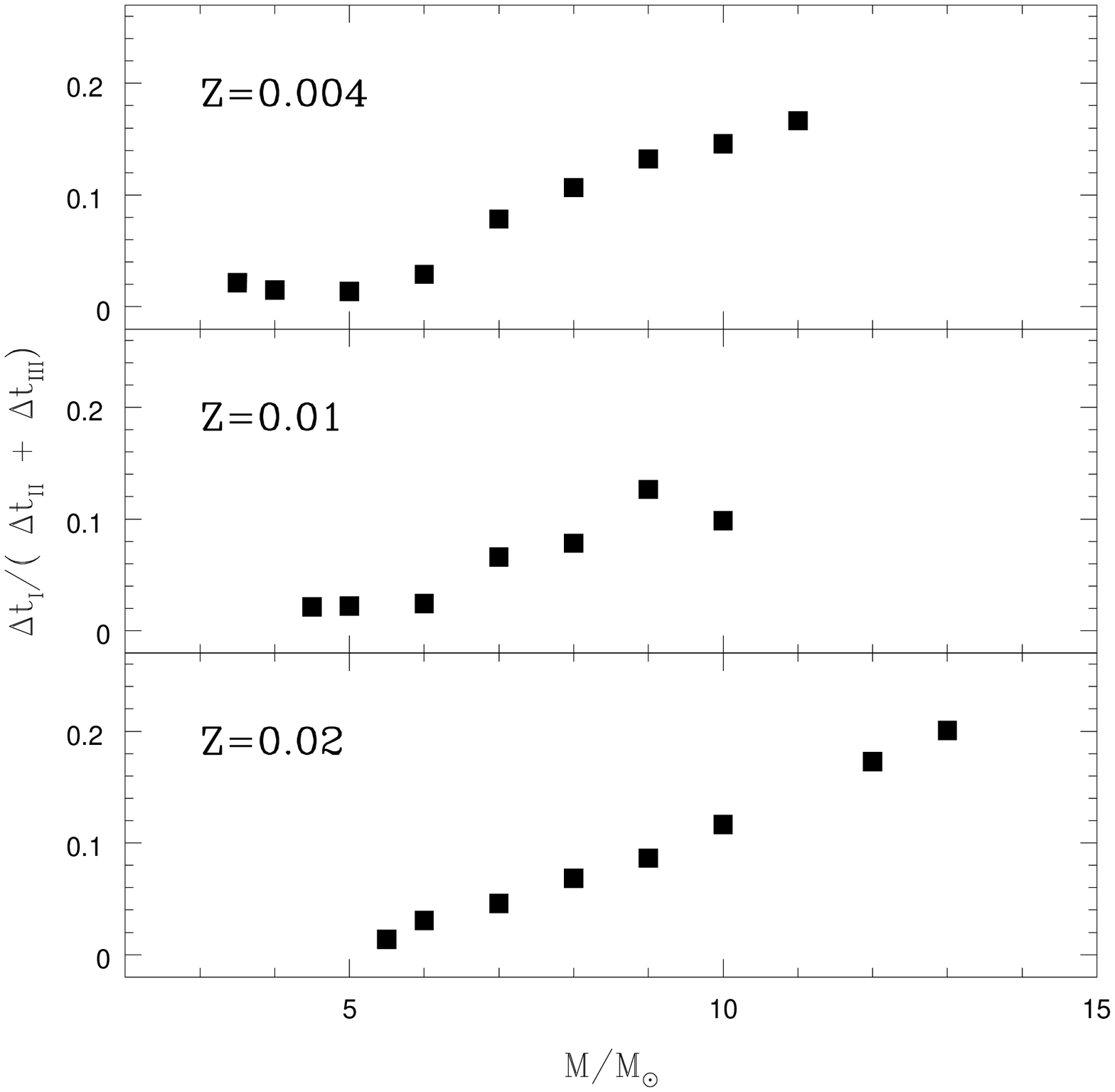,bbllx=0pt,bblly=0pt,bburx=590pt,bbury=720pt,width=12.5cm,
clip=,angle=0}
\vspace*{3pt}
\FigCap{
The ratios of duration of the first crossing to duration of the
II and III crossings for three metallicities calculated from models
presented by Bono {\it et al.} (2000).
}
\end{figure}

\begin{figure}[htb]
\hglue-0.5cm\psfig{figure=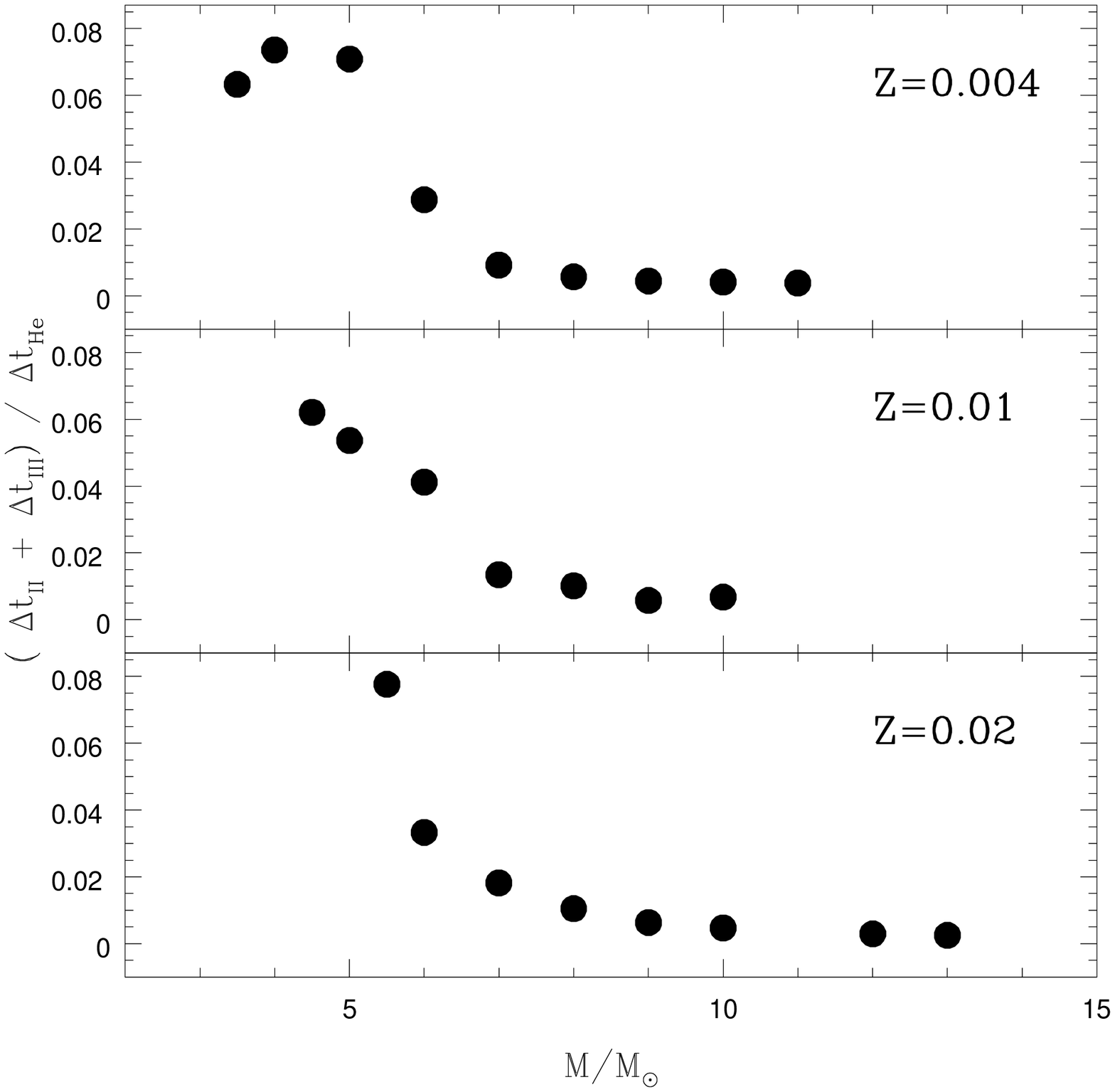,bbllx=0pt,bblly=0pt,bburx=590pt,bbury=720pt,width=12.5cm,
clip=,angle=0}
\vspace*{3pt}
\FigCap{
The ratios of duration of II and III crossings to duration of
the total core helium burning phase derived from Bono {\it et al.}
(2000) for stellar models with the tip of the blue loop at higher
effective temperatures than at the blue edge of the
instability strip.}
\end{figure}

\end{document}